\documentclass[pre,twocolumn]{revtex4}
\usepackage{psfrag}
\usepackage{amssymb,amsmath,amsthm}
\usepackage[dvips]{graphicx}
\usepackage{verbatim}
\usepackage[colorlinks=true,linkcolor=blue,citecolor=blue]{hyperref}

\begin{document}

\title{Arrays of stochastic oscillators: Nonlocal coupling, clustering, and wave formation}

\author{Daniel Escaff$^1$, Italo'Ivo Lima Dias Pinto$^2$, and Katja Lindenberg$^3$} 
\affiliation{ $^1$Complex Systems Group, Facultad de Ingenier\'{i}a y Ciencias
Aplicadas, Universidad de los Andes, Avenida Monse\~{n}or \'{A}lvaro del Portillo N$^{\text{o}}$ 12.455, Las Condes,
Santiago, Chile. \\$^2$Departamento de F\'{i}sica, CCEN, Universidade Federal da Para\'{i}ba,
Caixa Postal 5008, 58059-900, Jo\~{a}o Pessoa, Brazil \\
$^3$Department of Chemistry and Biochemistry and BioCircuits Institute,
University of California San Diego, La Jolla, California 92093-0340, USA
}

\begin{abstract}
\noindent We consider an array of units each of which can be in one of three states. Unidirectional transitions between these states are governed by Markovian rate processes. The interactions between units occur through a dependence of the transition rates of a unit on the states of the units with which it interacts. This coupling is nonlocal, that is, it is neither an all-to-all interaction (referred as global coupling), nor is it a nearest neighbor interaction (referred to as local coupling).The coupling is chosen so as to disfavor the crowding of interacting units in the same state.  As a result, there is no global synchronization. Instead, the resultant spatiotemporal configuration is one of clusters that move at a constant speed and that can be interpreted as \emph{traveling waves}. We develop a mean field theory to describe the cluster formation and analyze this model analytically. The predictions of the model are compared favorably with the results obtained by direct numerical simulations. 
\end{abstract}

\pacs{XXX }

\maketitle

\section{Introduction}

The emergence of self-organization in systems out of equilibrium has received a great deal of attention in the last few decades. These self-organizing systems have the fascinating common property that although formed from many microscopic constituents, they are capable of exhibiting coordinated macroscopic dynamics. Examples of this kind of behavior can be found in many contexts, ranging from spatial patterning \cite{PATT,Murray} to synchronization phenomena \cite{SYNC}. 

The interactions of the microscopic constituents in these systems range from global (all to all interactions), as in many models of synchronization~\cite{SYNC}, to local, as in many pattern forming systems where transport phenomena are diffusive~\cite{PATT}.  Between these two extremes lie what we call spatially nonlocal (but not global) interactions. For instance, neural models of pattern formation must take into account the effects of nonlocal interactions among neurons. Models that provide a good description of the mechanism involved in stripe formation in the visual cortex~\cite{Murray} involve firing rates of cells stimulated by close neighbors (activation) and depressed by more distant neighbors (inhibition). 

Nonlocal interactions are invoked in many contexts and may induce new dynamical states. A well known example is the appearance of chimera states~\cite{Chimera1} whose emergence seems to require nonlocal interactions. Chimeras have been found experimentally in the context of Belousov-Zhabotinksky chemical oscillators~\cite{Chimera3}.
The concept of nonlocal interactions has been generalized beyond spatial coupling. For instance, Abrams et al.~\cite{Chimera2} have suggested a model of a population of identical oscillators separated into two subgroups.  The oscillators in each subgroup are globally coupled to all other oscillators in that subgroup, and also to the oscillators of the other subgroup, but with different coupling strengths. The role of nonlocal interactions has also been pointed out in the context of vegetation dynamics in arid zones~\cite{Vegetation}, where scarcity of resources induces a variety of self-organizing patterns. Nonlocal interactions have also had an impact in the field of nonlinear optics~\cite{Optic}. In this case, nonlocality may arise, for instance,  in thermal nonlinear optical media and in left-handed materials. In the area of  population dynamics, nonlocal interactions can induce pattern formation \cite{PopulationPATT} and the stabilization of localized states \cite{PopulationLS}. In fact, when strong nonlocal coupling is taken into account \cite{Escaff01}, a new mechanism to stabilize localized states has recently  been reported \cite{Oto}, one that has no analog when the interactions are purely local. 

Most of the models mentioned above lead to self-organization in the mean field. They describe the self-organization process by a set of field variables  that account for densities related to the underlying microscopic dynamics. However, the mean field neglects the role that fluctuations play in the self-organization process, and may even miss some effects and phenomena entirely. For example, a model for population dynamics has been considered in Ref.~\cite{Cellular} (a binary cellular automaton) which exhibits pattern formation via intermittence that can not be described by a continuos mean field theory.

In this paper we focus on an array of units each of which can be in one of three possible states. Transitions between these states are governed by stochastic Markov processes. Hence, in contrast with the deterministic model presented in Ref.~\cite{Cellular}, here we can control fluctuations by a suitable scaling of the model parameters and implement a mean field theory for the self-organizing dynamics exhibited by the model. We fully analyze this mean field theory and compare the results with those obtained by direct numerical simulations of the model.

In the context of synchronization of coupled oscillators, units of discrete states may model a coarse-grained phase space of excitable and oscillatory units~\cite{Prager}. These kinds of systems have served as a fruitful tool to study synchronization as well as fluctuations \cite{Prager,Wood,Copelli,WoodCopelli,Escaff, Escaff2,Pinto}. In this paper we present a nonlocal generalization of the model proposed by Wood et al. \cite{Wood}, and analyze the resulting self-organized spatial structures. Originally, this model was proposed to study global synchronization, that is, to determine conditions for the majority of the units of the ensemble to oscillate together. For that reason, the coupling between units used in that study favors the ``crowding" or accumulation of units in the same state.  In contrast, here
we study the case of coupling that disfavors crowding. Due to the nonlocal interaction, this anti-crowding coupling may induce  the formation of traveling clusters of different phases of oscillation.

The goal of our work is to present a nonlocal generalization of the model of Wood et al.~\cite{Wood} to the anti-crowding regime.  We will show that this regime exhibits a novel self-organizing behavior, and we develop a mean field theory for this self-organizing phenomenon. We also discuss the capability of this mean field theory to describe what we observe from direct numerical simulations of the model. To fulfill these goals, the manuscript is organized as follows. In Sec.~\ref{sec2} we propose our model. In Sec.~\ref{sec3} we present a numerical study of the model, showing the novel self-organizing phenomenon. In Sec.~\ref{sec4} we fully analyze the mean field description of the process, and in Sec.~\ref{sec5}, we present our conclusions. Some mathematical derivations are presented in the Appendix.

\section{The Model}
\label{sec2}

\begin{figure}
\includegraphics[width =3.0in]{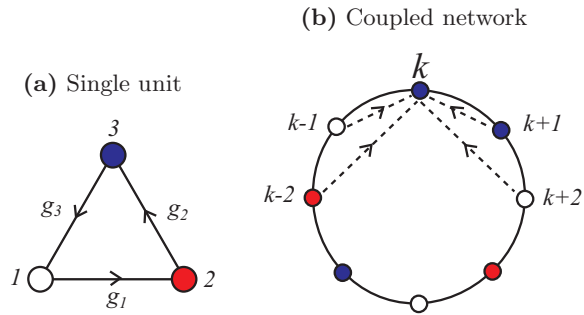}
\begin{picture}(0,0)
\put(-215,85){{\bf{(a)}} Single unit}
\put(-110,110){{\bf{(b)}} Coupled network}
\end{picture}
\caption{(Color online) (a) Single unit dynamics. Each colored dot represents a state. (b) Small coupled array of size $N=8$ with periodic boundary conditions. Each dot is a three-state unit, and its color represents the state of that unit. The unit at the top of the sketch ``interacts" with (or is ``aware" of) four neighbors.}
\label{fig01}
\end{figure}

We consider an array of $N$ units numbered $k=1,2,3,\ldots,N$, each of which may be in one of three possible states, say, state $1$ or state $2$ or
state  $3$. Transitions between these states are cyclical (unidirectional), from $1$ to $2$, $2$ to $3$, and $3$ to $1$, as illustrated in Fig.~\ref{fig01}(a). This unidirectionality of course implies that the system is out of equilibrium. The transtions are Markov processes of rates $g_i$, where $i$ is the initial state of the transition. The interaction among units is shown in cartoon form in Fig.~\ref{fig01}(b) and is modeled by taking the rate at which a transition occurs from one state to the next in a given three-state unit to depend on the states of a number of its neighbors. More precisely, suppose we focus on a particular unit, say unit $k$, and 
furthermore suppose that this unit  ``is aware" of the instantaneous states of $N_k$ of its neighbors. The transition rate out of state $i$ of that unit is then denoted by $g_{i}^k$ and is assumed to be of the Arrhenius form
\begin{equation}
g_{i}^{k}(t)=
\exp\left(a\,\frac{n_{i+1}^k(t) - n_{i}^k(t)}{N_k}\right).
\label{Rate1}
\end{equation} 
Here $n_{j}^k(t)$ is the number of units among the $N_k$ that are in state $j$ at time $t$, and $i+1=1$ when $i=3$. 

This model has been explored in considerable detail when the coupling parameter $a$ is positive~\cite{Wood}. When there are all to all interactions, that is, when $N_k=N$, the coupling is \emph{global}. The positive coupling parameter implies that if many units are in a given state, then they leave that state more slowly than if there are only a few. In the thermodynamic limit $N\to\infty$, the stable state of the system when the coupling parameter is small is the globally symmetric state, where $1/3$ of the units are in each state.  This state becomes unstable as the parameter $a$ increases beyond a critical value. There is then a supercritical transition to global synchronization~\cite{Wood}, and in the stable state most of the units oscillate around the three-state circuit in unison. These behaviors can be found via a mean field analysis.

Continuing with the global coupling case, the oscillations slow down as the parameter $a$ increases further. When the coupling parameter increases beyond a second critical value, the system undergoes another transition. The oscillatory state is lost via an infinite-period bifurcation, and the system reaches a static stationary state in which most of the units are in the same state.  There are of course three such possible over-crowded static states~\cite{Copelli}. In other words, as $a$ becomes more and more positive, and the tendency of crowding becomes more and more intense, the symmetry between the three states is broken.

\begin{figure*}
\includegraphics[width =7.0in]{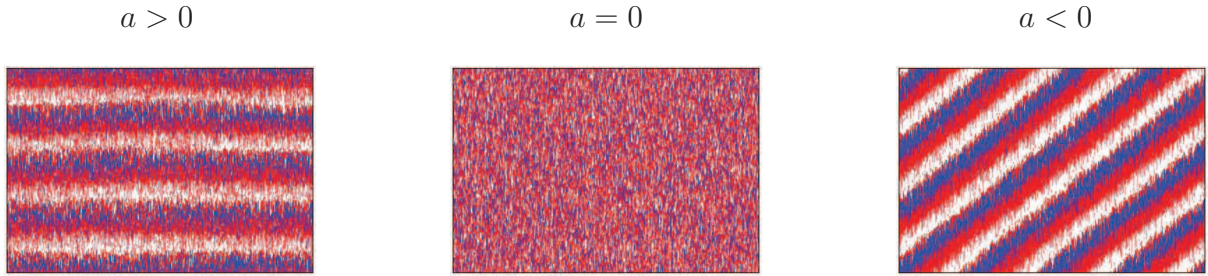}
\begin{picture}(0,0)
\put(-270,120){\large{$a=0$}}
\put(-440,120){\large{$a>0$}}
\put(-100,120){\large{$a<0$}}
%\put(-515,160){\large{\rotatebox{90}{Time}}}
\end{picture}
\caption{(Color online) Array of $N = 1024$ units, with $n = 200$. Left panel: $a=2$. Middle panel: $a=0$. Right panel: $a=-15$. In the three panels $t\in [0,40]$. Each color represents a different state as in Fig.~\ref{fig01}.} 
\label{fig02}
\end{figure*}

Generalizations of model (\ref{Rate1}) have also been considered~\cite{WoodCopelli}, enriching the bifurcation scenario. For instance, while the transition to synchronization obtained with global coupling and the transition rates given above is supercritical, a change in the structure of the exponent leads to a subcritical transition to synchronization. In any case, for all these variants of the model with global coupling, the crowding effect seems to be crucial for the occurrence of synchronous behavior.

At the opposite extreme of global coupling, but still with a positive coupling constant $a>0$, lies the case of \emph{local coupling}. In this case we have a regular network (e.g., hypercubic) in which each unit interacts only with its nearest neighbors~\cite{Wood}. Therefore, in 
Eq.~\eqref{Rate1} we now have $N_k=2d$, where $d$ is the spatial dimensionality of the array (global coupling is equivalent to infinite dimension). In this case, even in the thermodynamic limit global synchronization is not guaranteed.  Global synchronization requires a dimensionality $d\geq 3$, that is, a cubic or hypercubic array. Here it is no longer possible to use a mean field analysis. Instead, the transition to synchronization is found via a renormalization group analysis that requires numerical implementation~\cite{Wood}.

We now turn to the case of a negative coupling constant, $a<0$, which we call ``anti-crowding coupling." This form of coupling has been considered for a two-state model~\cite{Escaff}, where synchronization is not possible with either positive or negative coupling with Markovian transition rates.  We considered a model in which one of the transitions, say from state 1 to state 2,  is Markovian, but the reverse transition, from state 2 to state 1, is a non-Markovian process.  In this model the system memory is shortened when either state is overcrowded. When coupling is weak, the only steady state is quiescent, that is, on average the populations in states $1$ and $2$ remain static. 
However, at sufficiently strong coupling, the shortening of the memory when either state is overcrowded induces a high-amplitude oscillation that never destabilizes the quiescent state.  Anti-crowding coupling has not previously been considered for units of more than two states.

This is the case we consider in this paper, that is, anti-crowding coupling in arrays of three-state units with Markovian transition rates. It might be tempting to conjecture that anti-crowding coupling, $a<0$, might always induce disordered phases because the units attempt to differentiate one from another as much as possible. However, we will show that when \emph{nonlocal  coupling} is considered, the system can organize itself by forming propagating clusters. These clusters are spatially distributed, and, at a given location, they oscillate (alternate) between over-crowding and anti-crowding, which is another manifestation of propagating clusters. Thus, on average there are no globally ordered phases, and yet the system displays clear spatially structured synchrony.

To achieve this outcome, not only do we need to focus on negative coupling, but the model can not be entirely global nor entirely local, as were the models considered for positive coupling.  The array is still in the configuration shown in Fig.~\ref{fig01}(b), and each unit interacts with $n$ neighbors on each side. The transition rates are as given in Eq.~\eqref{Rate1}, with $N_k = 2n$ (since we take $N_k$ to be independent of $k$, all the units are identical). Global coupling corresponds to $n=N/2$, that is, $N_k=N$, and local coupling to $n=1$, that is, $N_k=2$.  We implement periodic boundary conditions as shown in the figure. The system dynamics are characterized by the three parameters $\left\{a,n,N\right\}$.

\section{Numerical observations}
\label{sec3}

In this section we display a numerical study of the model described at the end of the previous section for negative coupling constant, $a<0$, including a comparison of results with those of a positive coupling constant. We start by showing in Fig.~\ref{fig02} the typical scenarios that we observe in our simulations.  The horizontal axes represent positions (units) in the array, and the vertical axes show time. The size of our linear array is $N=1024$, and the number of neighbors with which any unit interacts is $n=200$ on each side. The different colors represent different states (1, 2 or 3). For sufficiently large positive values of $a$ (see Fig.~\ref{fig02}, left panel), we observe that our system oscillates as a whole from one overcrowded state to another, exhibiting global synchronization as occurs for the global coupling case in the thermodynamic limit~\cite{Wood}. For low positive coupling strength or no coupling at all (see Fig.~\ref{fig02}, middle panel), the system exhibits a completely disordered configuration where, on average, 1/3 of the units are in each state. 

On the other hand, and of interest to us here, when the coupling strength is large in magnitude and negative ($a<0$), (see Fig.~\ref{fig02}, right panel), the system exhibits a new form of self-ordering. Clusters appear in which one of the states is over-crowded, and yet an average over the full array shows no global crowding in any of the states.  The fuzyness at the edges of the fringes are caused by fluctuations that arise due to the Markovian transition rates and, more importantly,  due to the finite number $n$ of units coupled to each unit in the array. Note that the clusters move with a well-defined velocity. This appears as a well-defined slope in the spatiotemporal diagram. Due to the isotropy of the model, the motion of the clusters is equally likely to the left or right of the array. The direction depends on initial conditions and on fluctuations. Note that these clusters seem to be highly ordered in space, perhaps a reminiscence of some type of Turing self-organization.

%\onecolumngrid

\begin{figure*}
\includegraphics[width =6.0in]{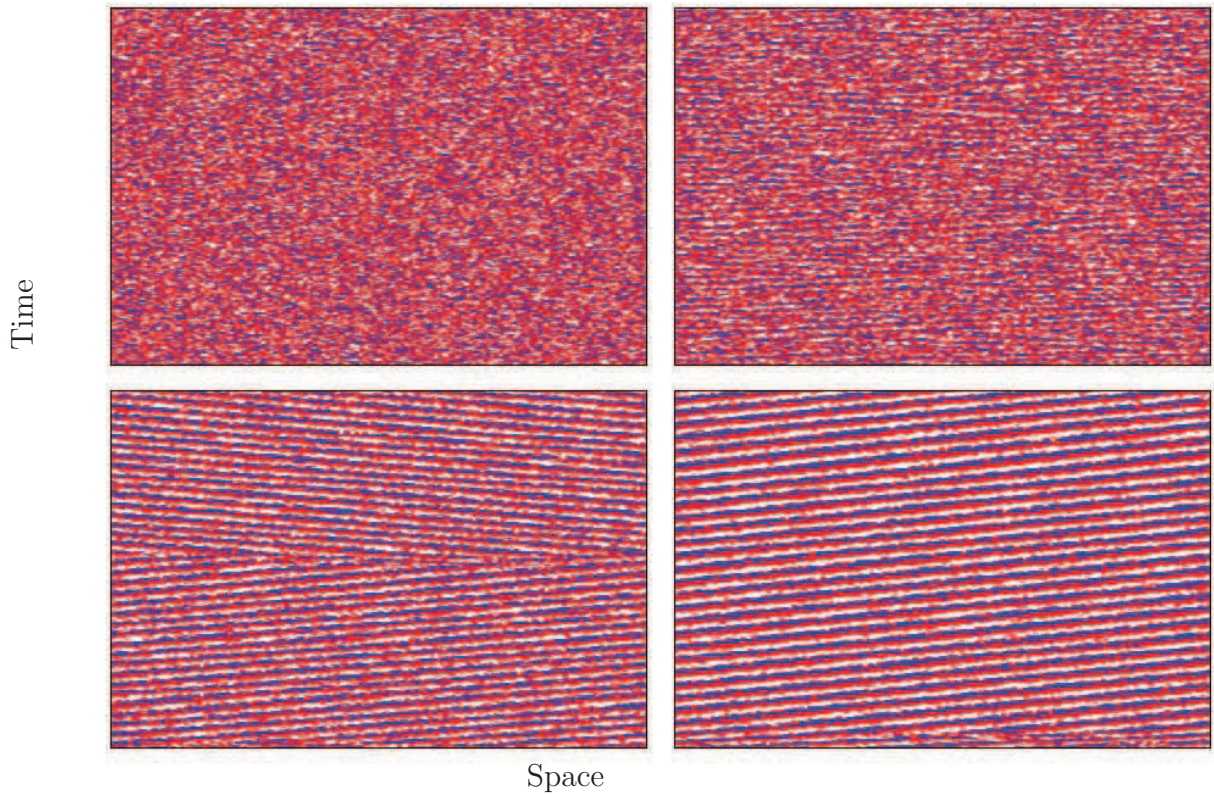}
\begin{picture}(0,0)
\put(-270,-5){\large{Space}}
\put(-465,160){\large{\rotatebox{90}{Time}}}
\end{picture}
\caption{(Color online) Array of $N = 512$ units with $n=104$. Top left panel: $a = -5$. Top right panel: $a = -9$. Bottom left panel: $a = -13$.  Top right panel:$a = -17$. In the four panels $t\in [0,400]$.}
\label{fig03}
\end{figure*}

%\twocolumngrid

To display the transition to self-organization more clearly, in Fig.~\ref{fig03} we follow the spatiotemporal configuration as the value of $a$ becomes increasingly negative.  Here we work with an array of $N = 512$ units and set $n = 104$. For $a = -5$ (upper left panel),  the spatiotemporal diagram is dominated by noise, and no pattern is evident. For $a = -9$  (upper right panel) we note the beginnings of the formation of clusters, but with a strong presence of noise. As we continue to increase the anti-crowding coupling strength, the clusters become increasingly evident ($a = -13$, lower left, and $a = -17$, lower right). We again observe that these clusters move with a well defined velocity. As noted earlier, the motion is equally likely to be in either direction, determined by the initial conditions and by the fluctuations. In fact, for $a = -17$, after a short transient, in this realization the clusters clearly move to the right. On the other hand, for $a = -13$, even though the clusters at first move to the right, they rather suddenly change  direction and continue to move to the left. This phenomenon is likely a fluctuation-induced transition. 

\begin{figure}
\includegraphics[width =1.65in]{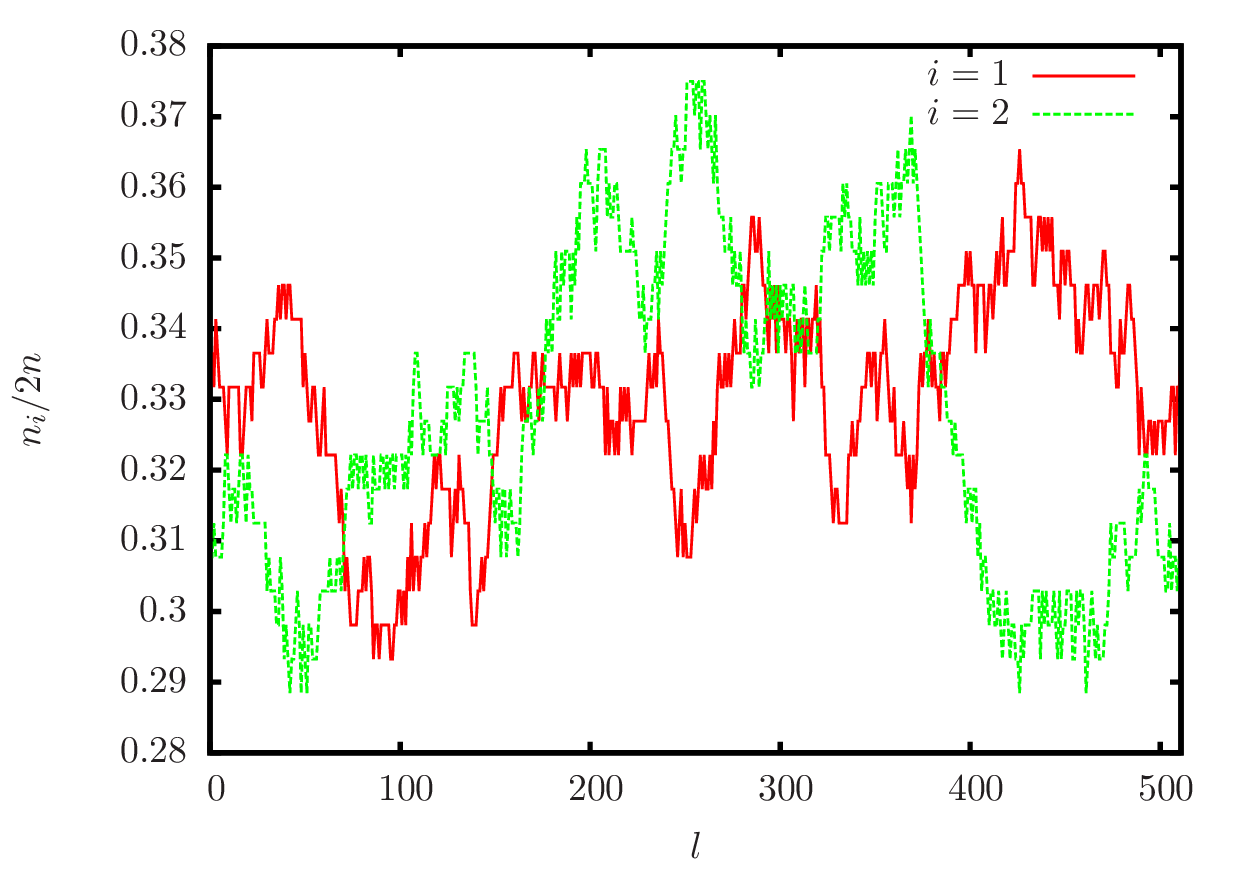}
\includegraphics[width =1.65in]{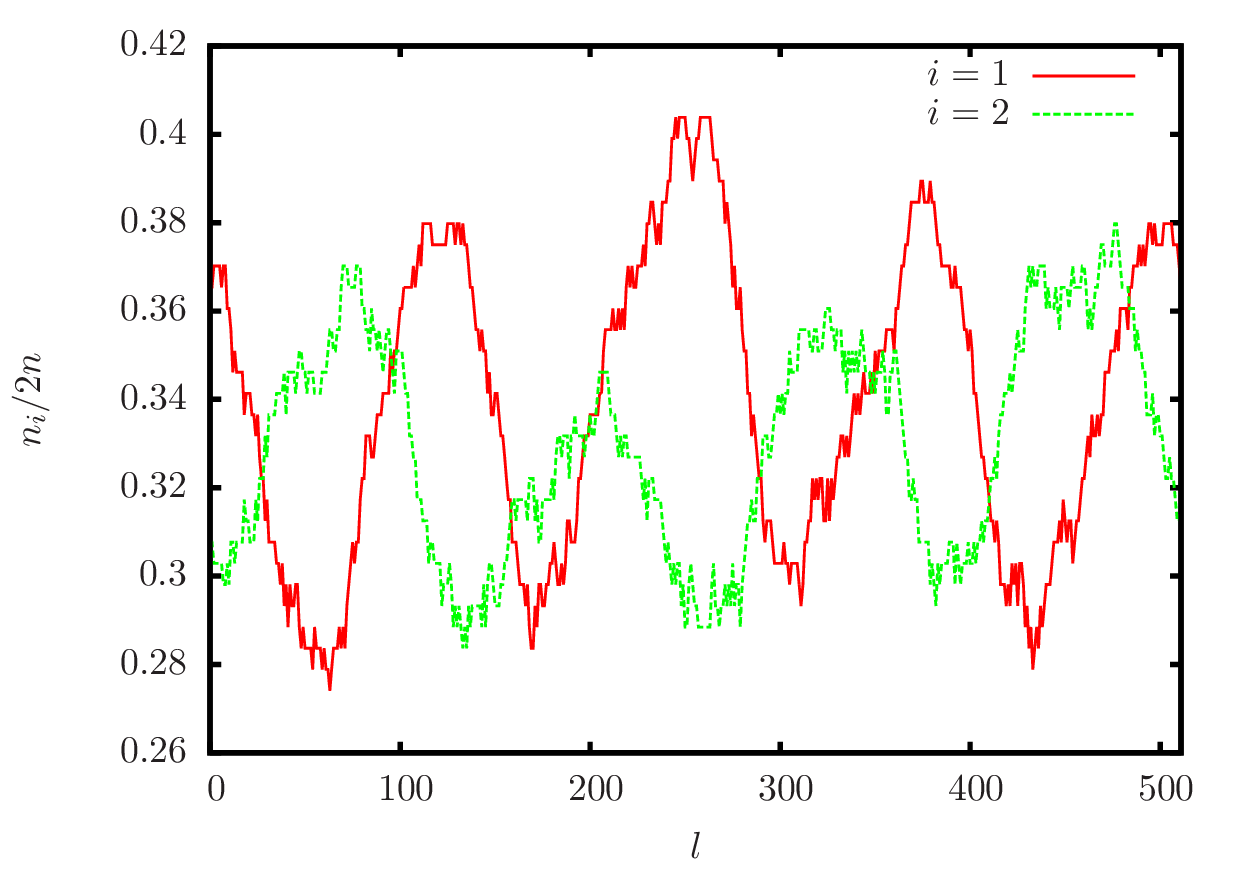}\\
\includegraphics[width =1.65in]{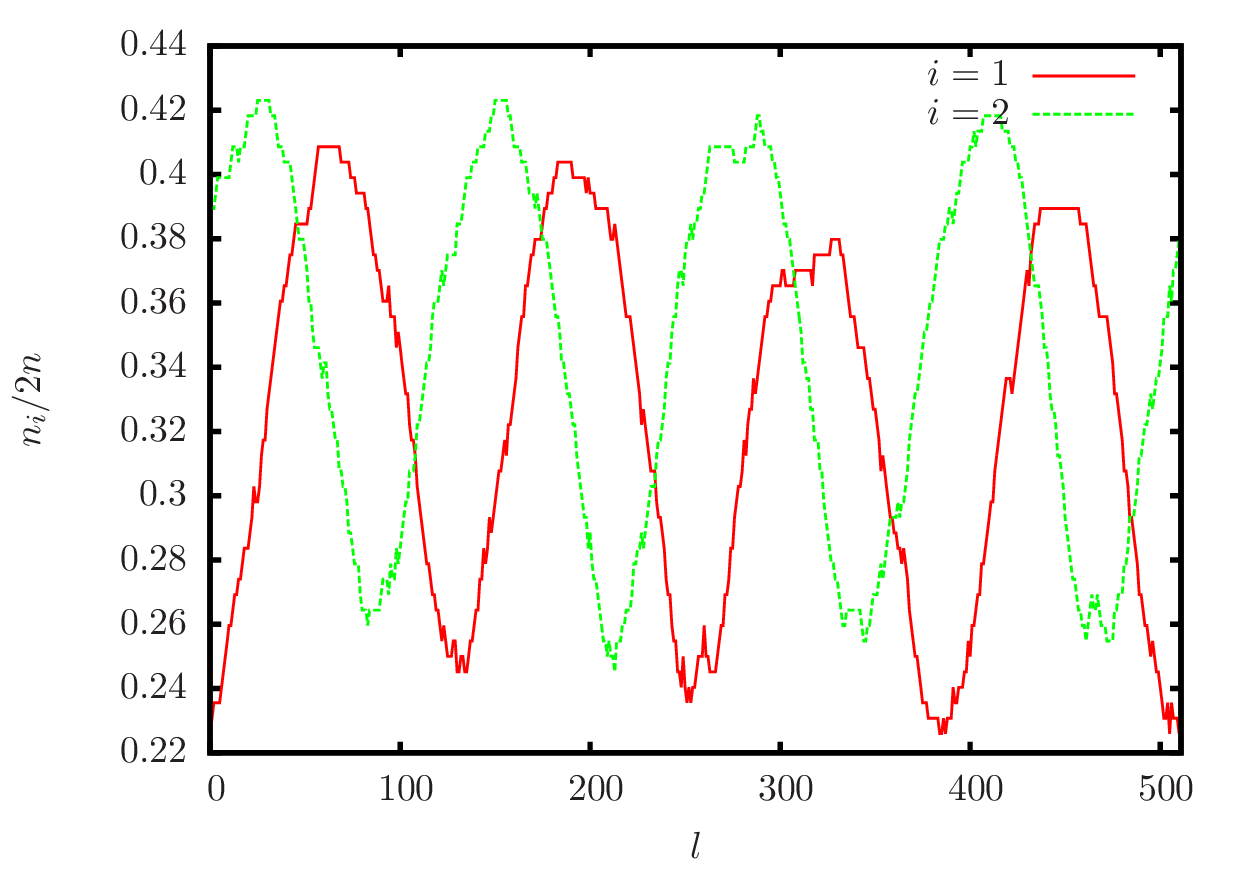}
\includegraphics[width =1.65in]{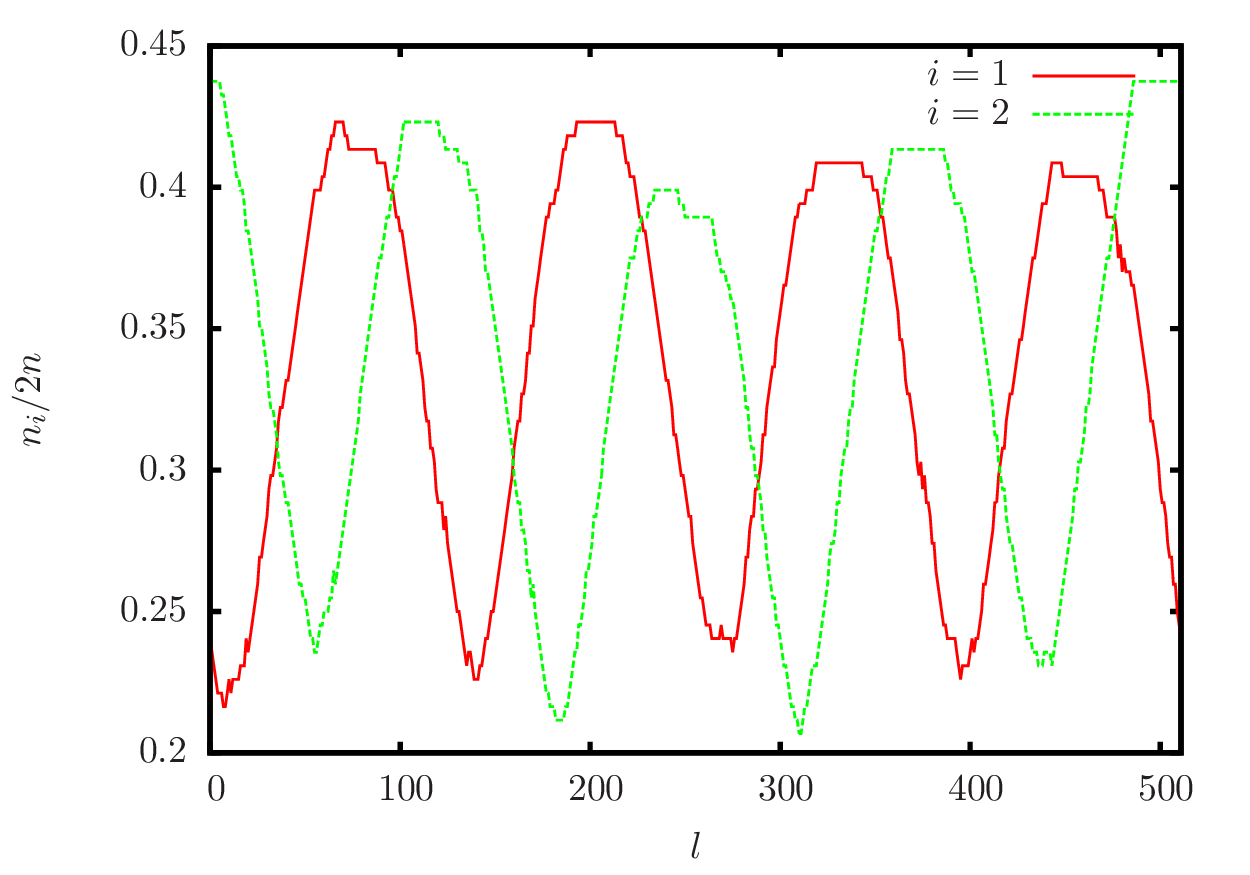}
\caption{(Color online) Spatial distribution of the number $n_i$ of interacting neighbours on both sides of a particular unit that are in state $i$ for different values of the coupling parameter. Red: state $i=1$. Green: state $i=2$. The parameters of the four panels are the same as in the four panels of Fig~\ref{fig03}.}
\label{fig04}
\end{figure}

\begin{figure}
\includegraphics[width =3.5in]{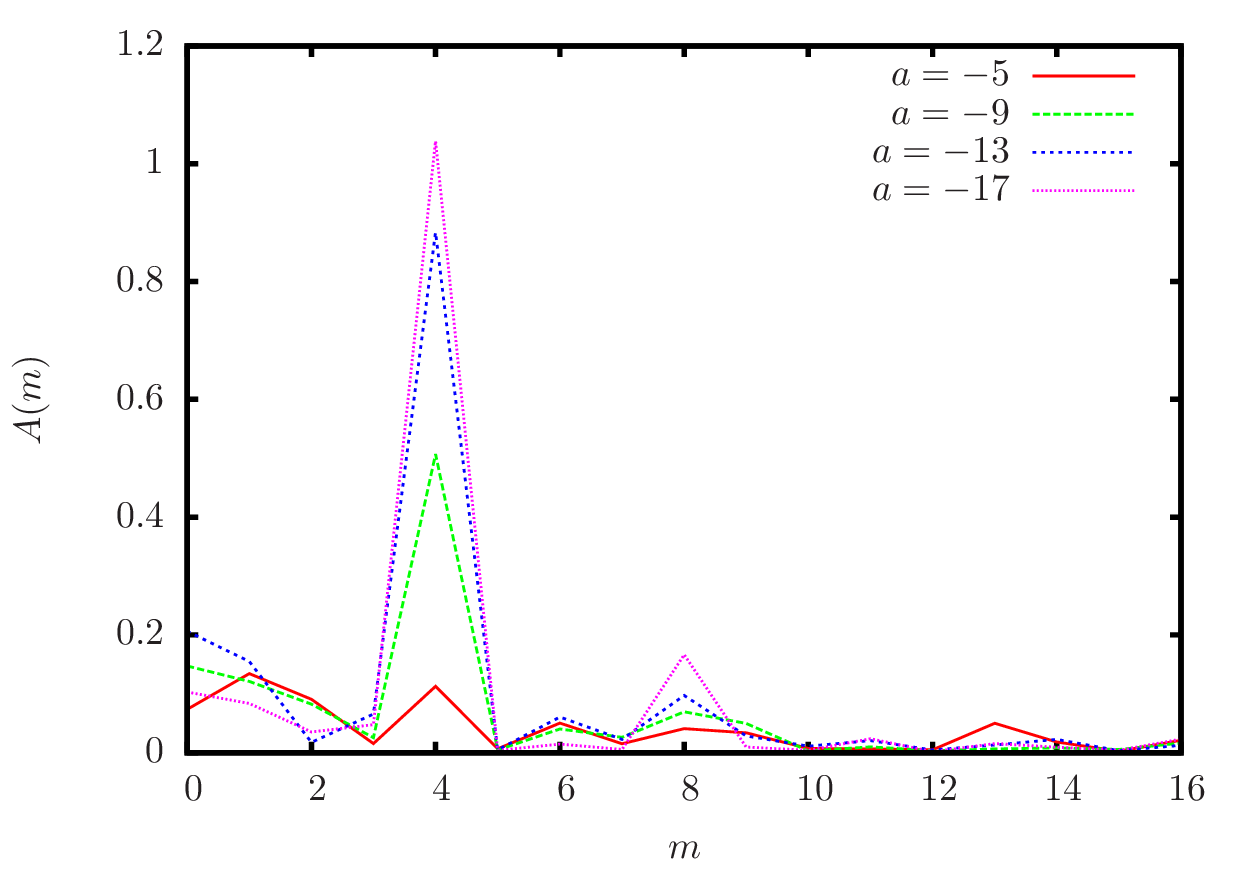}
\caption{(Color online) Absolute value of the Fourier transform of the spatial distribution of the $n_1$ neighbors of an arbitrary unit in state $1$ that interact with that unit and that are in state $1$.  The peak at $m=4$ is evident.
Different colors are used for different values of $a$ as indicated on the right upper corner: Red $a=-2.5$, green $a=-4.5$, blue $a=-6.5$, purple $a=-8.5$.}
\label{fig05}
\end{figure}

The periodic distribution of these clusters (affected by noise) can be seen if we define the quantity 
\begin{equation}
\nu_{i}^k(t) = \frac{n_{i}^k(t)}{2n},\label{den}
\end{equation}
which is less noisy than would be a rendition of the states of each unit along the array at a given time, especially as $-a$ increases. Figure~\ref{fig04} shows the spatial profile of this quantity for a given time for the same parameters as in Fig.~\ref{fig03}. As we increase the anti-crowding coupling strength, a clearer regular pattern emerges, showing that the system becomes more and more self-organized. Figure~\ref{fig05} displays the absolute value $A_m$ of the Fourier transform of $\nu_{1}^k(t)$, 
\begin{equation}
A_m=\frac{1}{\sqrt{N}} \left| \sum_{k=1}^{N} \nu_1^k e^{2\pi i mk/N}\right|
\end{equation}
(obviously, the $i$ in the exponent is the complex unit number, not a state index),
which exhibits a clear peak at wave number $m=4$. Therefore, we can describe these clusters as \emph{traveling waves} with well defined wave number $8\pi /N$, amplitude and speed. All of these phenomena occur in the presence of fluctuations.

\begin{figure}
\includegraphics[width =3.1in]{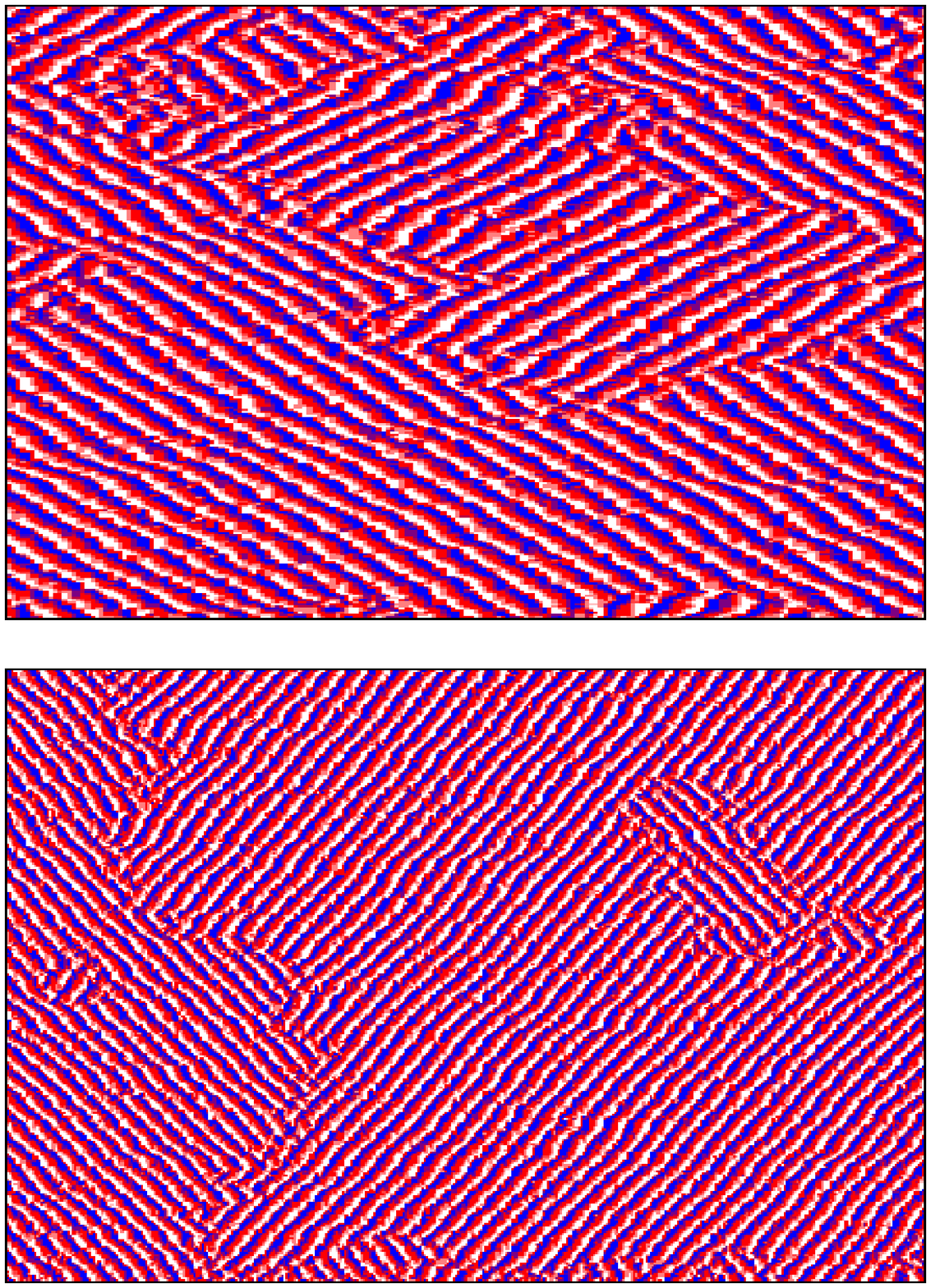}
\begin{picture}(0,0)
%\put(-270,-5){\large{Space}}
%\put(-515,160){\large{\rotatebox{90}{Time}}}
\end{picture}
\caption{(Color online) Upper panel: $N=200$, $n=7$, and $a=-16$. Bottom panel: $N=2000$, $n=9, and $ $a=-16$. In the lower panel we have plotted only the first 500 units of the array on the horizontal position axis to facilitate visualization of the wave patterns. In both panels $t\in [0,400]$.}
\label{fig07}
\end{figure}

Finally, in Fig.~\ref{fig07} we show the results of numerical simulations for two different values of $N$ and lower values of $n$. The effects of reducing this number are interesting.  The formation of traveling waves is again clear, a result we observe only if $n > 5$. For lower $n$ the system displays a noisy desynchronized phase. When the pattern appears in this regime as $n$ is modestly increased, however, one observes the coexistence of domains with left-moving and right-moving waves. These domains are separated by interphases that may be recognized as sources and sinks of waves. This is a well documented phenomenon for spatially extended oscillatory systems out of equilibrium in the mean field limit~\cite{Aranson}. However, contrary to those cases, in our case these defects seem to appear and move in a random way. In the top panel of Fig.~\ref{fig07} the defects seem to quickly appear and disappear. In the bottom panel, on the other hand, the defect trajectories are neater and longer. Their direction is quite random: perhaps the direction along the array executes a sort of Brownian motion. In addition, it is also possible to observe islands where one direction of propagation predominates immersed in a region where the other direction predominates.

\section{Mean field theory for wave formation}
\label{sec4}

We wish to support our numerical findings with analytic results.  We begin by constructing a master equation in discrete space which we then approximate by a continuous space master equation using a scaling argument. Next we implement a mean field approximation which leads us to a solution.  This solution in turn leads to a stable quiescent stationary state when the coupling parameter is small. As the parameter increases, in either the positive or negative directions, the quiescent solution becomes unstable. A positive coupling constant at this point then leads to a crowding solution that is spatially uniform but temporally oscillatory, as obtained in earlier work~\cite{Wood}.  A negative coupling constant at the point of instability leads to the onset of an anti-crowding spatio-temporal oscillatory solution, as we have seen in the previous section.  In this latter case we then go on to characterize the wave evolution when a single mode first becomes unstable, and analyze this self-organizing evolution with our numerical results in mind. 

\subsection{Master equation}

We start our analysis with the master equation that governs the evolution of the probability $p_{i}^k(t)$ that unit $k$ in our array of $N$ units is in state $i$ at time $t$. In view of the conservation of probabilities, $p_{1}^k(t)+p_{2}^k(t)+p_{2}^k(t)=1$, we only need equations for two of these probabilities: 
\begin{eqnarray}
\dot{p}_{1}^k(t) & = & g_3^k(t)p_3(t)-g_1^k(t)p_1(t)\nonumber\\
& = & g_{3}^{k}(t) - \left(g_{1}^{k}(t)+g_{3}^{k}(t)\right)p_{1}^k(t) 
\nonumber\\&&- g_{3}^{k}(t)p_{2}^k(t)
\label{master1} ,\\
\dot{p}_{2}^k(t) &=& -g_{2}^{k}(t)p_{2}^k(t) + g_{1}^{k}(t)p_{1}^k(t)
\label{master2},
\end{eqnarray}
where $g_{i}^{k}(t)$ is defined in Eq. (\ref{Rate1}).

\subsection{Continuous limit}

We next introduce the scaling variable
\begin{equation}
\nu=\frac{n}{N},
\end{equation}
and take the limits $N\rightarrow\infty$ and $n\to\infty$ keeping the value of $\nu$ constant. Space can then be described by the continuous variable $x=k/N$, with $x\in \left[0,1\right]$ and $dx=N^{-1}$. We implement the changes
\begin{equation}
p_{i}^k(t) \rightarrow  p_{i}(x,t), \text{ }\text{ } g_{i}^k(t) \rightarrow  g_{i}(x,t),
\end{equation}
so that the master equations (\ref{master1}) and (\ref{master2}) take the form
\begin{align}
\frac{\partial p_{1}(x,t)}{\partial t} =& g_{3}(x,t) - \left(g_{1}(x,t)+g_{3}(x,t)\right)p_{1}(x,t) \nonumber\\
&- g_{3}(x,t)p_{2}(x,t)
\label{masterCont1} ,\\
\frac{\partial p_{2}(x,t)}{\partial t} =& -g_{2}(x,t)p_{2}(x,t) + g_{1}(x,t)p_{1}(x,t)
\label{masterCont2}.
\end{align}
The periodic boundary condition takes the form 
\begin{equation}
p_{i}(x+1,t) = p_{i}(x,t).\label{CB}
\end{equation}

\subsection{The mean-field approximation}

Equations (\ref{masterCont1}) and (\ref{masterCont2}) are not autonomous unless we can express the rates $g_{i}(x,t)$ in terms of the probabilities $p_{1}(x,t)$ and $p_{2}(x,t)$. To do this, we go back to the definition $\nu_{i}^k(t)$ for the finite system given in Eq.~\eqref{den} and relate its statistical properties to the probabilities. Its time dependent mean value can be expressed as
\begin{equation}
\left\langle \nu_{i}^k(t)\right\rangle=\frac{1}{2n}\overset{n}{\underset{k'=1}{\sum}}\left(p_{i}^{k+k'}(t) 
+ p_{i}^{k-k'}(t) \right),
\end{equation}
and the standard deviation as
\begin{align}
&\sqrt{\left\langle \left(\nu_{i}^k(t) - \left\langle \nu_{i}^k(t)\right\rangle\right)^2 \right\rangle} =
\nonumber\\
& \frac{\sqrt{\overset{n}{\underset{k'=1}{\sum}}p_{i}^{k+k'}(t)-\left(p_{i}^{k+k'}(t)\right)^2
+ p_{i}^{k-k'}(t)-\left(p_{i}^{k-k'}(t)\right)^2}}{2n}.
\nonumber\\
\end{align}
Taking the limit $N\rightarrow \infty$, we have
\begin{equation}
\nu_{i}^k(t) \rightarrow \frac{1}{2\nu}\int_{-\nu}^\nu dx' p_{i}(x+x',t) + 
\mathcal{O}\left(\frac{1}{\sqrt{n}}\right).\label{MF}
\end{equation}
The last term, of $\mathcal{O}\left(1/\sqrt{n}\right)$, expresses the order of magnitude of the fluctuations. If $n$ goes to infinity with $N$, this mean field theory becomes exact. Otherwise, fluctuations will always be present, and this mean field theory will only describe the deterministic drifts that lead the system to self-organization.
Finally, in the continuous limit we set
\begin{equation}
g_{i}(x,t)= 
 \exp\left(\frac{a}{2\nu}\int_{-\nu}^\nu dx' \left(p_{i+1}(x+x',t) - p_{i}(x+x',t)\right)\right).
 \label{Rates11}
\end{equation}
With these substitutions, Eqs.~(\ref{masterCont1}) and (\ref{masterCont2}) become a closed deterministic dynamical system, that is, the equations become autonomous.

Note that for the global coupling case $\nu = 1/2$, Eqs.~(\ref{masterCont1}) and (\ref{masterCont2}) with \eqref{Rates11} can be reduced to the standard mean field equations presented in~\cite{Wood}. In fact, if we define the global probability
\begin{equation}
P_{i}(t)= \int_{0}^1 p_{i}(x,t)dx
\end{equation}
and apply the periodic boundary condition (\ref{CB}), the system (\ref{masterCont1}) and (\ref{masterCont2}) with \eqref{Rates11} can be reduced to the set of ordinary differential equations for $P_{i}(t)$ reported for the globally coupled network in~\cite{Wood}. We emphasize that this reduction is only possible if $\nu = 1/2$, which corresponds to the all to all interaction in this notation.

At the other extreme, we have the local coupling limit $\nu\rightarrow 0$, which is very singular. Moreover, since fluctuations decay as $1/\sqrt{n} = 1/\sqrt{\nu N}$,  the local limit is mostly ruled by fluctuations, and, therefore this mean-field description  fails. In this case, the renormalization group analysis reported in \cite{Wood}, which predicts that there is no synchronization in one dimension, is more appropriate. This is also consistent with our simulations for low $n$ ($n<5$), where even for strong coupling we observe no synchronization.

\subsection{Quiescent array and its linear stability}

\begin{figure}
\includegraphics[width =2.2 in]{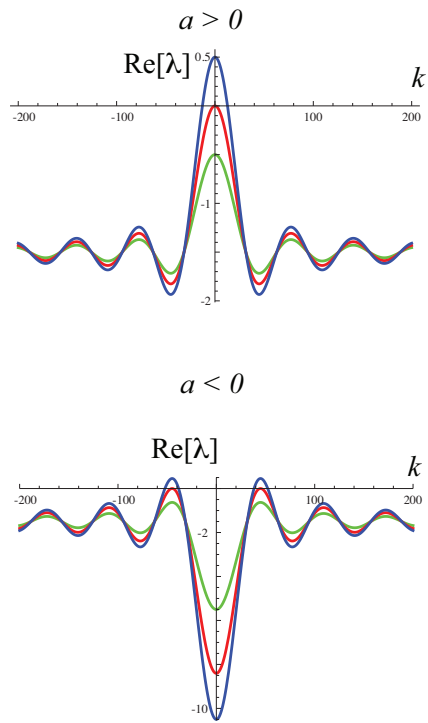}
\caption{(Color online) Spectrum of Eq.~(\ref{DisRel} for $\nu=0.1$. Top: green $a=1$; red $a=1.5$; blue $a=2$. Bottom: green $a=-4$; red $a=-6.905$; blue $a=-9$.}
\label{fig08}
\end{figure}

A trivial steady state solution of Eqs. (\ref{masterCont1}) and (\ref{masterCont2}) with Eq.~\eqref{Rates11} is the quiescent configuration
\begin{equation}
p_{1}(x,t) = p_{2}(x,t) = 1/3,\label{SymStates}
\end{equation}
which represents a disordered configuration in which each unit can be in any of the three possible states with equal probability. Each unit of course undergoes continual state changes, but on average the three states will be equally populated. The quiescent array is always a solution in the steady state because the system has permutation symmetry ($i\rightarrow i+1$, with $3\rightarrow 1$). However, the symmetry may be spontaneously broken by an instability of the symmetric configuration. More explicitly, suppose we perturb the symmetric configuration with a planar wave,
\begin{equation}
p_{j}(x,t)=1/3 + \varepsilon_j \exp\left(i\kappa x + \lambda t\right),
\end{equation}
with $\left|\varepsilon_j \right|\ll 1$. From (\ref{masterCont1}) and (\ref{masterCont2}) with \eqref{Rates11} we obtain the dispersion relation
\begin{equation}
\lambda(\kappa)= \frac{1}{2}\left(2\hat{a}(\kappa) - 3 \pm i\sqrt{3} \right),\label{DisRel}
\end{equation}
where
\begin{equation}
\hat{a}(\kappa)= a\frac{\sin\left(\kappa\nu\right)}{\kappa\nu}.
\label{Akappa}
\end{equation}
The quiescent configuration (\ref{SymStates}) becomes unstable when $Re\left[\lambda\right]$ is positive for some value(s) of $\kappa$.  For coupling that favors crowding ($a>0$), the instability  occurs at $a_c = 3/2$ with $\kappa=0$ regardless of the value of $\nu$.
With this zero wavevector one expects that at least near the onset of the instability the system as a whole begins to oscillate with no spatial structure. Farther from the transition value other wave vectors might lead to positive $Re\left[\lambda\right]$ nonzero values of $\kappa$, and spatial structures may appear.  In contrast with the crowding scenario, when anti-crowding coupling ($a<0$) is considered,
at the first point of instability we find that $\kappa\neq 0$, and therefore spatiotemporal patterning is expected. 
In this case the coupling at which the quiescent configuration first loses stability as well as the selected wave number depend on  $\nu$.
Fig.\ref{fig08} displays the shape of the spectrum (\ref{DisRel}) for both types of coupling. 

More can be said because we have not yet implemented the periodic boundary conditions (\ref{CB}), which restricts the allowed values of $\kappa$ to those that satisfy 
\begin{equation}
\kappa =\kappa_m= 2\pi m,
\label{Kappam}
\end{equation}
where $m$ is an integer. For a given value of $\nu$, the quiescent disordered solution destabilizes to an oscillatory solution of wave number $m_c$ when the parameter $a$ reaches the critical value $a_c$ that satisfies the condition 
\begin{equation}
 \frac{a_c\sin\left( 2\pi m_c\nu\right)}{ 2\pi m_c\nu}=\frac{3}{2}.
\label{CriticalPoint1} \\
\end{equation}
For all other values of $m$, 
\begin{equation}
 \frac{a_c\sin\left( 2\pi m \nu\right)}{ 2\pi m\nu}<\frac{3}{2}.
\label{CriticalPoint2} \\
\end{equation}

\begin{figure}[]
\begin{center}
\includegraphics[width =2.3in]{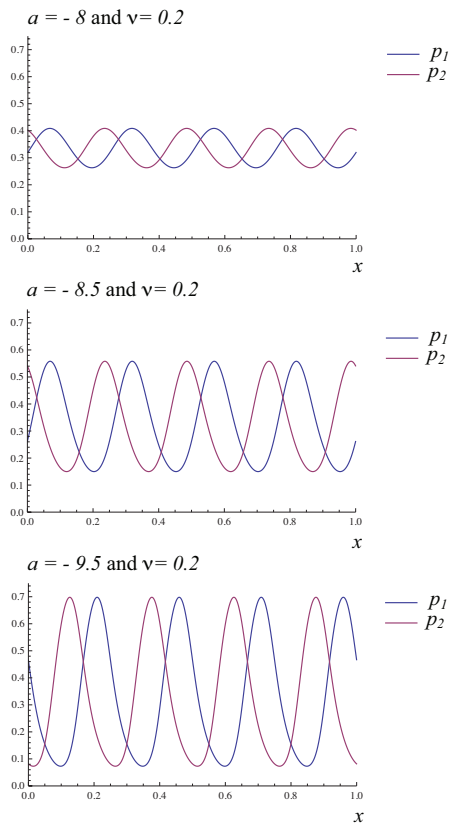}
\end{center}
\caption{(Color online) Probability distribution wave profiles at a given instant of time for different values of the coupling parameter beyond the critical value $a_c\cong-7.93$ for $\nu=0.2$, but within the range where only one wavenumber is unstable. Top panel: $a=-8.0$; middle panel: $a=-8.5$; lower panel: $a=-9.5$. Blue curves: $p_1$; purple curves: $p_2$. Note the increase in amplitude as the magnitude of the negative coupling parameter increases. } \label{fig09}
\end{figure}

As an example, let us fix $\nu=0.2$ and take $a$ as the control parameter. Then, the instability occurs at $a_c\cong-7.93$, selecting the wave number $m_c=4$. This wave number remains as the only unstable one up to $a\cong-9.62$, where $m=3$ also becomes unstable. Numerically solving the mean field equations (\ref{masterCont1}) and (\ref{masterCont2}) (using (\ref{Rates11})) in the range $a\in \left[-9.62,-7.93\right]$, we observe the formation of traveling waves with wave number $m=4$. The amplitude of these traveling waves increases as $a$ becomes more negative. Figure~\ref{fig09} displays the wave profile at a given instant for three different values of $a$, showing the increase in the amplitude.

\begin{figure}[]
\begin{center}
\includegraphics[width =2.7in]{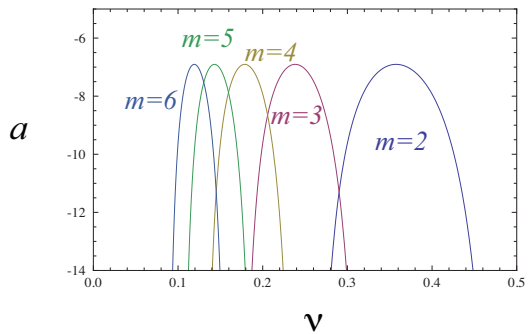}
\end{center}
\caption{(Color online) Regions of instability of modes $m\in\left\{2,...,6\right\}$ as a function of $a$ and $\nu$. Note that the first unstable mode in this figure at around $\nu = 0.2$ is $m=4$ at a value of $a_c \approx -7.93$ As $\nu$ increases, $m=3$ becomes unstable next, at a larger value of $a_c$. These results are consistent with those discussed in the text. Regions of overlap indicate that two modes are unstable.} \label{fig11}
\end{figure}

\begin{figure}[]
\begin{center}
\includegraphics[width =2.7in]{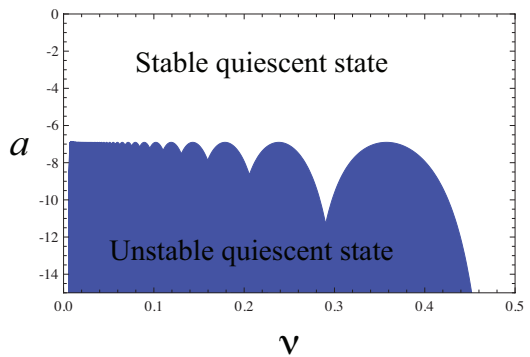}
\end{center}
\caption{(Color online) Regions of instability of a large range of modes, $m\in\left\{2,...,100\right\}$, as a function of $a$ and $\nu$. This figure covers almost the entire range of parameters and modes where the quiescent state is unstable for negative $a$.} 
\label{fig12}
\end{figure}

The wave number that first becomes unstable strongly depends on the values of $\nu$ and $a$. In Figs.~\ref{fig11} and \ref{fig12} we show the critical curves
\begin{equation}
Re\left[\lambda\left(2\pi m\right)\right] = 0
\end{equation}
for different values of the wave number $m$. Here $\lambda$ is given in Eq.~\eqref{DisRel}. Inside each tongue the associated mode is unstable. When tongues intersect there is more than 
one unstable mode.  Figure~\ref{fig11} displays the instability tongues encountered in a short interval of $m$ ranging from 2 to 6. Note that a wave with $m=1$ is not allowed for $\nu<1/2$. Figure~\ref{fig12} displays a large range of $\nu$ where the quiescent configuration is unstable, and waves are seen with $m$ ranging from 2 to 100. This covers almost the entire regime where the quiescent state is unstable. Tongues begin to appear for lower values of $\nu$ as $m$ increases, and begin to leak into the region $\nu\rightarrow 0$ where the mean field approximation is not valid.

We have thus shown that in the mean field approximation we are able to calculate the critical coupling constant for a given value of the range of the interactions at which the quiescent configuration of the array with equal populations in each of the three states first becomes unstable. We are also able to calculate the wave number of the first oscillatory instability, the next one that follows when the coupling strengrh increases, etc. The results obtained in the mean field approximation mimic those obtained from direct numerical simulations of the microscopic model carried out in the previous section. The wave number predicted by the mean field theory as the first instability agrees with our results from the microscopic dynamics. The critical coupling predicted by the mean field theory will be discussed further  below.

\subsection{Single mode wave evolution}

In order to characterize the wave evolution just beyond the first appearance of the instability, we consider the case when a single mode $\kappa_c$ is unstable. Then, we use the Ansatz to write the probability just beyond the appearance of the first instability as 
\begin{align}
p_j \left(x,t\right) \approx & 1/3 + \sqrt{\varepsilon}\psi_j   \left(A_L\left(\tau\right)\exp\left(i\kappa_c x + \Omega t\right)\right.
\nonumber\\&
\left. + A_R\left(\tau\right)\exp\left(-i\kappa_c x + \Omega t\right)
\right) + c.c.
\label{Ansatz1} ,
\end{align}
where $c.c.$ stands for complex conjugate, $A_L$ is the amplitude of the left propagating waves and $A_R$ is the amplitude of the right propagating waves. 

The constant $\varepsilon$ that appears in Eq.~(\ref{Ansatz1}) is related to the distance from threshold,
\begin{equation}
\varepsilon = \frac{\left(a - a_c\right)\sin\left( 2\pi m_c\nu_c\right)}{ 2\pi m_c\nu_c},
\label{Varepsilon}
\end{equation}
which is assumed to be small and positive, $0<\varepsilon\ll 1$ (note that the sin function here is negative, cf. Eq.~\eqref{CriticalPoint1}). The oscillatory term is thus a small ``distance" away from the uniform solution.
$\Omega$ is the natural frequency of oscillation of the system at the first instability,
\begin{equation}
\Omega = \frac{\sqrt{3}}{2} = Im\left[\lambda\right]. 
\end{equation}
We implement the usual assumption of perturbation theories, namely, that the amplitudes of the oscillations vary much more slowly than the oscillations themselves.  This is captured in the amplitude dependence on a slow time scale $\tau$,
\begin{equation}
\tau=\varepsilon t.  
\end{equation}
The remaining constants are given by  $\psi_1 = 1- i\sqrt{3}$ and $\psi_2 = -2$.

Detailed calculations for arriving at evolution equations for the amplitudes are given in the Appendix; here we  summarize and analyze the resulting equations, which are:
\begin{align}
\frac{\partial A_L}{\partial \tau} =& A_L - \left(\alpha\left|A_L\right|^2 + \beta\left|A_R\right|^2\right) A_L,
\label{AmplitudeLeft} \\
\frac{\partial A_R}{\partial \tau} =& A_R - \left(\alpha\left|A_R\right|^2 + \beta\left|A_L\right|^2\right) A_R
\label{AmplitudeRight},
\end{align}
where
\begin{align}
\alpha =& 54 - \frac{27\left(2 + i\sqrt{3}\right)}{2\left(1 + i\sqrt{3}-\cos\left(k_c\nu_c\right)\right)},
\nonumber\\
\beta =& 108 - \frac{81\left(2 + i\sqrt{3}\right)}{3 i\sqrt{3}- 2a_c +3}.
%\nonumber.
\end{align}
To analyze equations (\ref{AmplitudeLeft}) and (\ref{AmplitudeRight}), we separate the real and imaginary parts,
\begin{align}
A_L = \rho_L\exp\left(i\theta_L\right)\quad&\quad A_R = \rho_R\exp\left(i\theta_R\right),
\nonumber\\
\alpha =\alpha_{Re}+i\alpha_{Im}\quad&\quad\beta =\beta_{Re}+i\beta_{Im}.
%\nonumber.
\end{align}
Therefore, the moduli of the amplitudes satisfy an independent set of equations,
\begin{align}
\frac{\partial \rho_L}{\partial \tau} =& \rho_L - \left(\alpha_{Re} \rho_L^2 + \beta_{Re} \rho_R^2\right) \rho_L,
\label{RHOAmplitudeLeft} \\
\frac{\partial \rho_R}{\partial \tau} =& \rho_R - \left(\alpha_{Re} \rho_R^2 + \beta_{Re} \rho_L^2\right) \rho_R,
\label{RHOAmplitudeRight}
\end{align}
while the evolution of the phases is completely determined by the moduli,
\begin{align}
\frac{\partial \theta_L}{\partial \tau} &=-\alpha_{Im} \rho_L^2 - \beta_{Im} \rho_R^2,
\label{THETAAmplitudeLeft} \\
\frac{\partial \theta_R}{\partial \tau} &=-\alpha_{Im} \rho_R^2 - \beta_{Im} \rho_L^2.
\label{THETAAmplitudeRight}
\end{align}

Note that the evolution of the moduli is generated by the potential
\begin{align}
\mathcal{U}\left(\rho_L, \rho_R\right)=&-\frac{1}{2} \left(\rho_L^2 + \rho_R^2\right)+\frac{\alpha_{Re}}{4} \left(\rho_L^4 + \rho_R^4\right)
\nonumber\\&
+\frac{\beta_{Re}}{2}\rho_L^2\rho_R^2,\label{Potential}
\end{align}
in terms of which we can write
\begin{equation}
\frac{\partial \rho_L}{\partial \tau} = -\frac{\partial \mathcal{U}}{\partial \rho_L} \qquad \quad \frac{\partial \rho_R}{\partial \tau} = -\frac{\partial \mathcal{U}}{\partial \rho_R}.
\end{equation}
Hence, the dynamics of the amplitudes $A_L$ and $A_R$ are obtained directly from the  minimization of the potential (\ref{Potential}), in fact, from the evolution equation of the potential via the condition
\begin{equation}
\frac{d \mathcal{U}}{d\tau}= -\left( \left(\frac{\partial \mathcal{U}}{\partial \rho_L}\right)^2 + \left(\frac{\partial \mathcal{U}}{\partial \rho_R}\right)^2\right)\leq 0.
\end{equation}

\begin{figure}[]
\begin{center}
\includegraphics[width =2.7in]{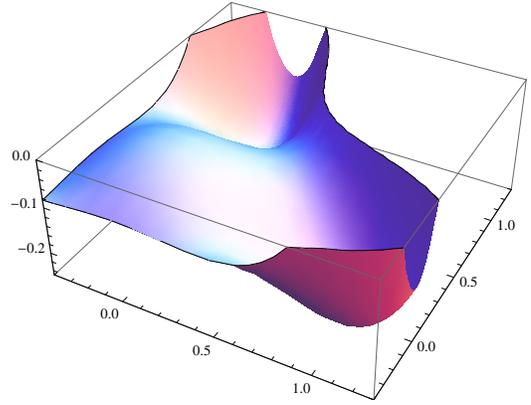}
\end{center}
\caption{(Color online) A typical rendition of the potential surface $\mathcal{U}$ as a function of $\rho_L$ and $\rho_R$.} \label{fig13}
\end{figure}

Figure~\ref{fig13} shows the typical shape of the potential (\ref{Potential}). It has a maximum at $A_L = A_R =0$, which gives the quiescent configuration $p_1=p_2=1/3$. Since we have assumed $\varepsilon>0$ in our calculation, the quiescent configuration appears as an unstable fixed point in this analysis. We have two minima that represent traveling waves,
\begin{align}
&\rho_L = 0 \qquad \quad  \rho_R = \frac{1}{\sqrt{\alpha_{Re}}},
%\nonumber
\\
&\rho_L = \frac{1}{\sqrt{\alpha_{Re}}} \qquad \quad \rho_R = 0,
%\nonumber
\end{align}
which are completely symmetric due to the isotropy of the model. 
There is also a saddle point which represents a standing wave,
\begin{equation}
\rho_L = \rho_R = \frac{1}{\sqrt{\alpha_{Re} + \beta_{Re}}}.
\end{equation}

\begin{figure}[]
\begin{center}
\includegraphics[width =2.1in]{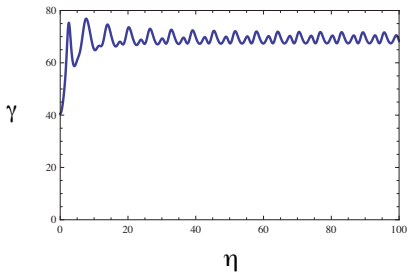}
\end{center}
\caption{(Color online) Typical plot of $\gamma$ vs $\eta$. } \label{fig14}
\end{figure}

To establish which of these solutions is stable, we compute the eigenvalues associated with each. For the standing wave the eigenvalues are
\begin{equation}
\Lambda_1 = -2\qquad \quad  \Lambda_2 = \frac{\beta_{Re} - \alpha_{Re}}{\alpha_{Re} + \beta_{Re}},
\end{equation}
and for the traveling waves
\begin{equation}
\Lambda_1 = -2\qquad \quad  \Lambda_2 = \frac{ \alpha_{Re}-\beta_{Re}}{\alpha_{Re}}.
\end{equation}
Therefore, if $\alpha_{Re}<\beta_{Re}$, the traveling wave is an attractor and the standing wave a hyperbolic point. The reverse inequality implies that  the attractor corresponds to the standing wave. To establish the direction of the inequality we compute the quantity
%\begin{equation}
$\gamma\left(\eta\right) =\beta_{Re} - \alpha_{Re}$,
%\end{equation}
where $\eta = 2\pi m_c\nu$ is the only relevant variable since $a_c$ can be written in terms of $\eta$ using the critical relation (\ref{CriticalPoint1}). Figure~\ref{fig14} displays the function $\gamma\left(\eta\right)$, showing that it is always positive, i.e.,  the traveling waves are always stable, while the standing wave is always unstable. This is consistent with the fact that, in direct numerical simulations of the microscopic dynamics, we always observe traveling waves (we have never seen a standing wave).

Therefore, as the quiescent configuration loses its stability, traveling waves begin to form, selecting a direction of propagation determined by the initial condition and, when the number of units is finite, by the intrinsic fluctuations of the system.  Moreover, the noise may induce switching in the direction of propagation, as we see, for example, in the bottom left panel of Fig.~\ref{fig02}. That is, driven by fluctuations the system jumps between the two minima of the potential (\ref{Potential}).

At the mean field level, the system is attracted by the stable fixed points of Eqs.~(\ref{RHOAmplitudeLeft}), (\ref{RHOAmplitudeRight}), (\ref{THETAAmplitudeLeft}) and (\ref{THETAAmplitudeRight}), that is,
\begin{equation}
p_j\left(x,t\right) \cong 1/3 + \mathcal{A}\cos\left(\kappa_c\left(x \pm vt\right) + \theta_j\right)
+ \mathcal{O}\left(\varepsilon\right). \label{WaveApprox}
\end{equation}
Here $\kappa_c = 2\pi m_c$, $\mathcal{O}$ accounts for higher order corrections, the amplitude is given by
\begin{equation}
\mathcal{A}= 4\sqrt{\frac{\varepsilon}{\alpha_{Re}}},\label{Amplitude}
\end{equation}
the phase velocity takes the form
\begin{equation}
v = \frac{\Omega - \varepsilon(\alpha_{Im}/\alpha_{Re})}{\kappa_c} + \mathcal{O}\left(\varepsilon^2\right),\label{Velocity}
\end{equation}
and the phase shift is
\begin{equation}
\theta_2 - \theta_1 = \pm\frac{2\pi}{3}.
\label{PhaseShift}
\end{equation}
The  $+$ sign gives the phase shift for the left-wave, and the $-$ for the right-wave.

\subsection{Comparison between analytical predictions and direct numerical simulations}

To assess how well this theory captures the microscopic rules that govern this system, we compare the results of the theory with numerical simulations of the amplitude and the phase shift.
%We do so for the three quantities predicted by the theory: the amplitude, the phase velocity, and the phase shift.

We can not compare numerical simulations directly with the amplitude in Eq.~\eqref{Amplitude} because the probabilities $p_j(x,t)$ are not directly related to any observable density.  The measured density is $\nu_i^k(t)$ defined in Eq.~\eqref{den}. The two quantities are related by Eq.~\eqref{MF}, 
\begin{equation}
\nu_{j}^k(t) \approx \nu_{j}(x,t) = \frac{1}{2\nu}\int_{-\nu}^\nu dx' p_{j}(x+x',t),
\end{equation}
where $x=k/N$. With Eq.~\eqref{WaveApprox} we then have 
\begin{equation}
\nu_j\left(x,t\right) \approx 1/3 + \mathcal{B}\cos\left(\kappa_c\left(x \pm vt\right) + \theta_j\right),
\end{equation}
where
\begin{equation}
\mathcal{B} = \left(\frac{\sin\left(\kappa_c\nu\right)}{\kappa_c\nu}\right)\mathcal{A}.
\label{NewAmplitude}
\end{equation}
Only the amplitude must thus be modified for direct comparison; the phase and phase velocity remain unchanged.

\begin{figure}[]
\begin{center}
\includegraphics[width =3.0in]{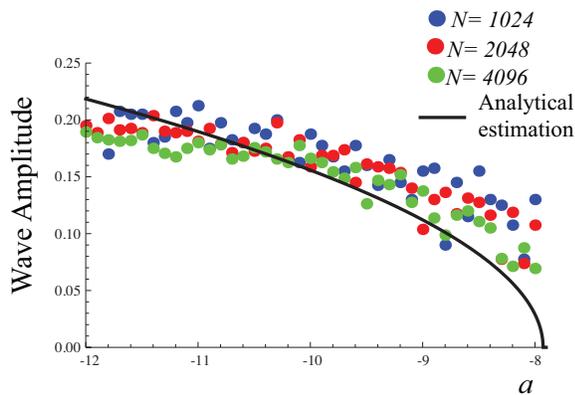}
\end{center}
\caption{(Color online) Solid line: Amplitude $2\mathcal{B}$ from Eq.~\eqref{NewAmplitude}. Dots are numerical simulations for three different array sizes, as indicated in the figure. } 
\label{fig:Amplitude}
\end{figure}

The comparison of this analytical prediction and the numerical simulations for $\nu=0.2$ and three values of array size $N$ are shown in Fig.~\ref{fig:Amplitude}. Numerically, we compute the average difference between the maximum and minimum of the signal $\nu_{j}^k$, which corresponds to $2\mathcal{B}$ in the mean field approach. The analytical prediction is clearly quite good.  The numerical simulation results are insensitive to the array size; the scatter may in part be due to the fluctuations that the mean field theory does not capture.  Note that the simulated amplitude does not go to zero at the critical coupling predicted by the mean field theory. Perhaps this is the realization of a well documented phenomenon in pattern forming systems (both experimentally and theoretically), that in the presence of noise the pattern appears below onset without noise. This is known as a \emph{noisy precursor} \cite{Agez}, or a \emph{stochastic Turing pattern} \cite{Butler}.

Note also that the coupling strengths explored in Fig.~\ref{fig:Amplitude} go beyond the range of instability of a single mode. Actually, for $\nu=0.2$ and $a<-9.62$, the modes with $m=4$ and $m=3$ are both linearly unstable. However, the amplitude equation only considers the first unstable mode, in this case $m=4$ (which destabilizes at $a_c=-7.93$). In spite of this, the predictions seem to be correct (at least in order of magnitude) even far from the regime of strict applicability of the perturbation assumption. Perhaps in the nonlinear saturation processs the mode $m=3$ remains inactive, even up to $a=-12$.

\begin{figure}[]
\begin{center}
\includegraphics[width =2.1in,angle=-90]{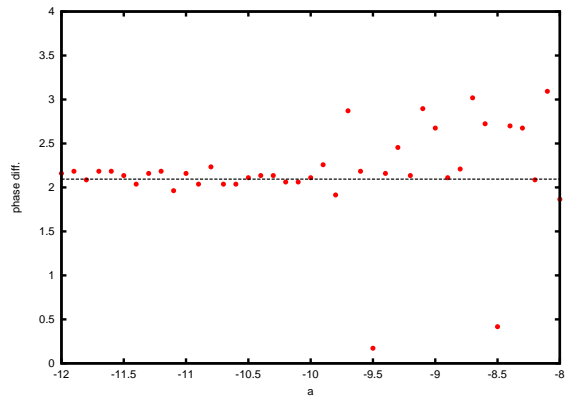}
\end{center}
\caption{(Color online)Phase shift.  Solid line: Eq.~\eqref{PhaseShift}.  Dots: numerical simulations. The large scatter of the simulations at the larger values of the phase shift may be due to fluctuations that become more pronounced near the critical point. }
\label{fig:phasediff}
\end{figure}

%The other two predicted quantities are the phase velocity and the phase shift. 
%The analytic results Eqs.~\eqref{Velocity} and \eqref{PhaseShift} along with the results of the numerical simulations for these quantities are shown 
In Fig. \ref{fig:phasediff} we show the analytic result Eq.~\eqref{PhaseShift} along with the results of the numerical simulation of this quantity. The agreement is good, although there is scatter in the numerical simulations, especially at larsge values of the coupling parameter.
Perhaps here, again, the fluctuations become more relevant near the critical point.

%\begin{figure}[]
%\begin{center}
%\includegraphics[width =2.1in,angle=-90]{Fig16.eps}
%\end{center}
%\caption{Phase velocity. \textcolor{red}{\bf WE NEED THE ANALYTIC RESULT OF Eq.~\eqref{Velocity}, IN BLACK PLEASE}. Dots: numerical simulations.} 
%\label{fig:phasevelocity}
%\end{figure}

%\subsection{Multi-modes waves}

%\begin{figure}[]
%\begin{center}
%\includegraphics[width =3.3in]{MM-MF.eps}
%\end{center}
%\caption{ $p_1(x,t)$, $\nu=0.002$ and $a=-8$} \label{fig14}
%\end{figure}

\section{Conclusion and future prospects}
\label{sec5}

In this work we have analyzed arrays of three-state units nonlinearly coupled to one another.  Transitions between states in each unit are uni-directional (that is, this is an out-of-equilibrium driven system), and the transition rates are Markovian stochastic processes.  This stochasticity makes this an inherently noisy system.  The coupling is modeled by further assuming that the Markovian transition rates of each unit depend on the states of the units to which each unit is said to be coupled. The model is inspired by one first studied by Wood et al.~\cite{Wood}.  In that work the coupling is taken to favor crowding, that is, a unit is likely to remain longer in a state occupied by a larger number of the neighbors to which it is coupled.  That coupling leads to synchronization of the entire system via a supercritical bifurcation provided the coupling is sufficiently strong and the system is at least three-dimensional. Extensive analysis of many aspects of that model was carried out, especially in the case of global coupling. 

The most important new feature of the work in this paper is that the sign of the coupling coefficient has been changed from positive to negative.  This means that the coupling favors anti-crowding, that is, a unit is likely to leave a state more rapidly if it is coupled to many other units in that state.
We have  taken the coupling to be non-local,
that is, it is neither nearest neighbor nor global.  Instead, each unit is coupled to a finite fraction $\nu=n/N$ of the total number of units $N$, and this fraction is one of the parameters of the system.  The most noteworthy result of the model is a transition to the formation of clusters that alternate between different oscillation phases and that propagate in space.  The transition from a disordered configuration to the formation of clusters occurs when the coupling coefficient is sufficiently negative. How negative this coupling must be depends on the fraction of units coupled to each unit.  

We presented results of numerical simulations of the equations of motion and compared them favorably with those that result from a mean field theory that yields analytic results. We calculated the conditions for instability of the quiescent behavior of the array (the bifurcation here us supercritical), and observed regions of instability of different frequency modes as a function of the coupling fraction. We were able to present an analytic treatment of the wave evolution just beyond the first appearance of the instability. The analytic prediction of the instability and the wave evolution just beyond agree semiquantitatively with those of the simulations.
Quantitative differences between the two are few and arise because the mean field theory does not address the fluctuations that are captured by the simulations when $N$ is finite. 

This work can be extended in many new directions. First, we analyzed regions of instability where a single mode is unstable.  The analysis can be extended to regions where two or more modes of different frequencies are simultaneously unstable. Furthermore, by multi-mode analysis one may be able to capture the defect dynamics observed for low $\nu$ (see, e.g., Fig.~\ref{fig07}). We can also consider different forms of coupling, in particular, a coupling considered by Wood et al.~\cite{WoodCopelli} that leads to a subcritical bifurcation in the case of positive coupling constant. It would be interesting to consider the effects of spatial diversity caused by the presence of more than one type of unit in the array.  It would also be interesting to consider an array of units in which the symmetry of the three states is broken by having different transition rates between different states; here we have considered these three rates to be equal. We have also not systematically studied the consequences of additional fluctuations introduced into the system by having short arrays.  In simulations shown in this paper we have seen the important role of fluctuations in choosing an initial direction of propagation and, in some cases, a sudden reversal of that direction.  We have made considerable progress in a number of these extensions of this work.

Finally, we mention an avenue of work that will take us back to the case of positive coupling.  In the work first reported by Wood et al.~\cite{Wood}, the system was analyzed very carefully in the determination of the value of the coupling constant leading to synchronization and the frequency of the oscillations at that point as a function of all the system parameters.  It was noted that with the coupling originally used in that work, the frequency of oscillations decreases with increasing coupling, but no further note was taken of this result. Later, Assis et al.~\cite{Copelli} took this result further and noted that a continued increase in the coupling constant led to a symmetry-breaking second transition, where the units slowed down completely and the majority of units simply remained motionless in one of the three states.  We have continued in this positive direction and increased the coupling constant even further, and have noted the appearance of interesting moving and fluctuating patches of units in different states.  We are continuing this analysis as well.

\bigskip

\section*{Acknowledgments}
D. E. thanks the financial support of FONDECYT Project No. 1140128. I. L. D. P. thanks the CNPq and CAPES for financial support.  K. L. thanks the support of the NSF under Grant No. PHY-0855471.

\bigskip

\bigskip

%\section*{APPENDIX}
\appendix*
\section{Single mode amplitude equations}
\label{appendix}

In order to derive the amplitude equations \eqref{AmplitudeLeft} and \eqref{AmplitudeRight}, we first introduce the shifted distributions
\begin{equation}
q_{j}(x,t) = p_{j}(x,t) -1/3,
\end{equation}
and define the vector
\begin{equation}
\vec{q} = 
\left(
\begin{array}
[c]{ll}%
q_1 \\
q_2 \\
\end{array}
\right),
\end{equation}
which sets the quiescent configuration at $\vec{q} = 0$.
Equations~\eqref{masterCont1} and \eqref{masterCont2} with the mean field approximation \eqref{Rates11} then take the form
\begin{equation}
\frac{\partial \vec{q}}{\partial t} = \mathbb{L}\vec{q} + N\left(\vec{q}\right),\label{EvEq}
\end{equation}
where we have separated the linear part $\mathbb{L}\vec{q}$ from the nonlinear part $N\left(\vec{q}\right)$. 

We can write the nonlinear portion explicitly as a Taylor series expansion,
\begin{equation}
N\left(\vec{q}\right) = \sum_{l=2}^{\infty}N_{l}\left(\vec{q}\right),
\end{equation}
where $N_{l}\left(\vec{q}\right)$ denotes products of powers of $q_1$ and $q_2$ of total order $l$. In other words, for any constant number $\xi$, 
\begin{equation}
N_{l}\left(\xi\vec{q}\right)=\xi^l N_{l}\left(\vec{q}\right).
\end{equation}
The terms in this series can be evaluated by expanding Eqs.~\eqref{masterCont1}, \eqref{masterCont2}, and \eqref{Rates11} in Taylor series.

The linear operator $\mathbb{L}$ is given by
\begin{equation}
\mathbb{L} = 
\left(
\begin{array}
[c]{ll}%
a\mathcal{L}-2 & \quad -1\\
1 & \quad a\mathcal{L}-1\\
\end{array}
\right),
\end{equation}
where we have introduced the operator
\begin{equation}
\mathcal{L}f(x) = \frac{1}{2\nu}\int_{-\nu}^\nu dx' f(x+x').
\end{equation}
Note that the spatial translational invariance of the operator $\mathcal{L}$ means that it is diagonal in Fourier space,
\begin{equation}
\mathcal{L}\exp\left(i\kappa x\right) = \left(\frac{\sin\left(\kappa \nu\right)}{\kappa \nu}\right)\exp\left(i\kappa x\right).
\end{equation}

Due to the periodic boundary condition 
\begin{equation}
\vec{q}(x,t)=\vec{q}(x+1,t),
\end{equation}
we can expand the shifted distribution in terms of Fourier modes,
\begin{equation}
\vec{q}(x,t)=\overset{\infty}{\underset{m=-\infty}{\sum}}
 \vec{\phi}_m (t)
\exp\left(i\kappa_m x\right),
\end{equation}
where $\kappa_m$ is defined in Eq.~\eqref{Kappam},
and $\vec{\phi}_m (t) = \vec{\phi}_{-m} (t)$ because $\vec{q}(x,t)$ is real.

\subsubsection{Critical point}

If we linearize the evolution equation (\ref{EvEq}) around the quiescent state $\vec{q} = 0$, we obtain the set of equations
\begin{equation}
\dot{\vec{\phi}}_m = L^{(m)}\vec{\phi}_m.
\end{equation}
Here
\begin{equation}
L^{(m)} = 
\left(
\begin{array}
[c]{ll}%
\hat{a}\left(\kappa_m\right)-2 \quad& -1\\
1 \quad& \hat{a}\left(\kappa_m\right)-1\\
\end{array}
\right),
\end{equation}
where $\hat{a}\left(\kappa\right)$ is defined in Eq.~\eqref{Akappa}. 

From these equations we can easily deduce the critical conditions Eqs.~\eqref{CriticalPoint1} and \eqref{CriticalPoint2}. 
Moreover, at the critical point $a=a_c$, $m=m_c$ for a given $\nu$,
\begin{equation}
L^{(m_c)} =  L_0 =
\left(
\begin{array}
[c]{ll}%
-1/2 \quad &  -1\\
1 \quad & 1/2\\
\end{array}
\right),
\end{equation}
which has the eigenvalues and eigenvectors 
\begin{equation*}
L_0\vec{\psi} = i\Omega\vec{\psi}  ~~\text{and}~~ L_0\vec{\psi}^* = -i\Omega\vec{\psi}^*,
\end{equation*}
where $\Omega=\sqrt{3}/2$ is the natural frequency of the system, and
\begin{equation}
\vec{\psi} = 
\left(
\begin{array}
[c]{ll}%
\psi_1 \\
\psi_2\\
\end{array}
\right)=
\left(
\begin{array}
[c]{ll}%
1 - i\sqrt{3} \\
-2\\
\end{array}
\right).
\end{equation}

\subsubsection{Unfolding the critical point}

Next we investigate the nonlinear saturation of the instability. Toward this purpose, we unfold the critical point,
\begin{equation}
a = a_c + \delta a,
\end{equation}
and define the expansion parameter
\begin{equation}
\varepsilon = \frac{\delta a\sin\left( 2\pi m_c\nu\right)}{ 2\pi m_c\nu}
\end{equation}
(cf. Eq.~\eqref{Varepsilon}).
Since $\varepsilon >0$ this corresponds to an expansion into the unstable situation. Moreover, near onset of the instability $\varepsilon \ll 1$.

Note that, we are only moving the control parameter $a$, keeping the chosen value of $\nu$ fixed. Furthermore, we are addressing the generic situation of penetrating only one of the tongues of Fig.~\ref{fig11} or \ref{fig12}, not two at the same time. Therefore, we fix an appropriate value of $\nu$, which fixes $m_c$ while we  move $a$. 

In this situation, we have a four-dimensional critical sub-space $\mathbb{S}$ generated by the vectors that belong to the basis
 \begin{align}
\mathcal{S} =& \left\{\vec{\psi}\exp\left(i\kappa_c x + i\Omega t\right),\vec{\psi}\exp\left(-i\kappa_c x + i\Omega t\right),\right.
\nonumber\\&
\left.\vec{\psi}^{*}\exp\left(-i\kappa_c x-  i\Omega t\right),\vec{\psi}^{*}\exp\left(i\kappa_c x - i\Omega t\right)\right\}
\nonumber,
\end{align}
that is, the critical sub-space $\mathbb{S}$ corresponds to the set of all possible linear combinations of the elements that belong to the basis set $\mathcal{S}$.

Hence, we can write the evolution equation (\ref{EvEq}) in the form
\begin{equation}
\frac{\partial \vec{q}}{\partial t} = \left(\mathbb{L}_0 + \varepsilon\mathbb{L}_1\right)\vec{q} +  \sum_{l=2}^{\infty}N_{l}\left(\vec{q}\right),\label{EvEq2}
\end{equation}
where, we have separated the linear operator into two parts, namely, the critical part
\begin{equation}
\mathbb{L}_0 = 
\left(
\begin{array}
[c]{ll}%
a_c~\mathcal{L}-2 \quad & -1\\
1 \quad & a_c~\mathcal{L}-1\\
\end{array}
\right),
\end{equation}
and the unfolding part, which has the form
\begin{equation}
\mathbb{L}_1 = 
\left(
\begin{array}
[c]{ll}%
1 \quad & 0\\
0 \quad & 1\\
\end{array}
\right)\left(\frac{\kappa_c \nu_c}{\sin\left(\kappa_c \nu_c\right)}\right)\mathcal{L}.
\end{equation}

Now we introduce the assumptions that we use in our perturbative calculations. We work under the following hypotheses:
\bigskip
\begin{enumerate}
\item \textbf{Hypothesis 1:} We can expand the solution of the evolution equation in a power series in $\sqrt{\varepsilon}$, at least near the onset of the instability.
\item \textbf{Hypothesis 2:} The first order of the expansion is completely determined by the critical modes that belong to the critical sub-space $\mathbb{S}$.
\item \textbf{Hypothesis 3:} There are two time scales.

\textbf{A fast time scale,} which is related to the oscillation of frequency 
\begin{equation}
\Omega = \frac{\sqrt{3}}{2}.
\end{equation}

\textbf{A slow time scale,} which is related to the growth of the unstable modes, 
\begin{equation}
\tau = \varepsilon t.
\end{equation}

We formally treat these two time scales as independent variables by setting
\begin{equation}
\frac{\partial }{\partial t} \rightarrow \frac{\partial }{\partial t} + \varepsilon\frac{\partial }{\partial \tau}.
\end{equation}

\end{enumerate}

We next introduce the Ansatz
\begin{align}
\vec{q}(x,t,\tau)= & \sqrt{\varepsilon} \left\{\vec{\psi} A_L\left(\tau\right)\exp\left(i\kappa_c x + \Omega t\right) + \right.
\nonumber\\&
\left.\vec{\psi} A_R\left(\tau\right)\exp\left(-i\kappa_c x + \Omega t\right) + c.c \right\}
\nonumber\\&
+\overset{\infty}{\underset{\alpha=2}{\sum}} \varepsilon^{\alpha/2}\vec{W}_\alpha (x,t,\tau)
\label{Ansatz}.
\end{align}
Substituting this Ansatz into the evolution equation (\ref{EvEq2}), and separating each order $\varepsilon^{\alpha/2}$, we obtain a set of equations of the form
\begin{equation}
\left(\mathbb{L}_0 - \frac{\partial }{\partial t}\right)W_\alpha= \mathcal{F}_\alpha ~~\text{with}~~ \alpha\in\left\{2,...,\infty\right\}.
\end{equation}
The right hand sides $\mathcal{F}_\alpha$ of this set of equations must be computed order by order from Eq.(\ref{EvEq2}). Furthermore, each $\mathcal{F}_\alpha$ depends on the results for previous orders. 

We have therefore transformed the nonlinear evolution equation (\ref{EvEq2}) into an infinite set of linear inhomogeneous equations for the corrections $W_\alpha$. All of these equations involve the same linear operator $\left(\mathbb{L}_0 - \frac{\partial }{\partial t}\right)$. Hence, the validity of the expansion demands that all the right hand sides $\mathcal{F}_\alpha$ belong to the image of the operator $\left(\mathbb{L}_0 - \frac{\partial }{\partial t}\right)$.

Since we are formally treating the two time scales $(t,\tau)$ as independent variables, from the Ansatz (\ref{Ansatz}) it follows that the solutions $W_\alpha$ will be periodic functions of $t$ with a period $T=2\pi/\Omega$. More precisely, 
\begin{equation}
W_\alpha(x,t,\tau) = W_\alpha(x,t + T,\tau) = W_\alpha(x + 1,t,\tau).
\end{equation}
Therefore, we can generate these functions from the basis
\begin{equation}
\mathcal{B} =\left\{\Theta_{mn},\Theta_{mn}^*\right\}_{m,~n = -\infty}^{\infty} ,
\end{equation}
where
\begin{equation}
\Theta_{mn} = \vec{\psi}\exp\left(i2\pi m x + in\Omega t\right).
\end{equation}
That is, $W_\alpha$ and $\mathcal{F}_\alpha$ belong to the space $\mathbb{B}$ of the all possible linear combinations of the elements of the basis set $\mathcal{B}$. Moreover, for any linear combination of the elements of this basis we assume that the coefficients can be functions of the slow time scale $\tau$.

Note that the basis of the critical sub-space $\mathcal{S}\subset\mathcal{B}$, and it may be written in this notation as
\begin{equation}
\mathcal{S} = \left\{\Theta_{m_c 1}, \Theta_{-m_c 1},\Theta_{m_c 1}^*, \Theta_{-m_c 1}^*\right\}.
\end{equation}
Furthermore, the critical sub-space $\mathbb{S}$ corresponds to the kernel of the operator $\left(\mathbb{L}_0 - \frac{\partial }{\partial t}\right)$, that is,
\begin{equation}
\text{if}~\sigma\in\mathbb{S} \qquad \mbox{then} \qquad \left(\mathbb{L}_0 - \frac{\partial }{\partial t}\right)\sigma =0.
\end{equation}
We can use this fact to elucidate the image of this operator. We define the complementary set
\begin{equation}
\mathcal{B}^C = \mathcal{B} - \mathcal{S} ,
\end{equation}
and denote the sub-space of all possible linear combinations of the elements of $\mathcal{B}^C$ by $\mathbb{B}^C$. It is then clear that the image of the operator $\left(\mathbb{L}_0 - \frac{\partial }{\partial t}\right)$ corresponds to $\mathbb{B}^C$.

Therefore, at each order we must impose the solvability condition
\begin{equation}
\mathcal{F}_\alpha\in\mathbb{B}^C .\label{SolCon}
\end{equation}
Moreover, to avoid ambiguities in the selection of the particular solutions for the corrections $W_\alpha$, we also impose $W_\alpha\in\mathbb{B}^C$. This choice can be motivated as in perturbation theory in quantum mechanics. The first order in the perturbative expansion (\ref{Ansatz}) belongs to the critical sub-space $\mathbb{S}$. We are therefore requiring that the highter orders have, in some sense, no projection in $\mathbb{S}$. This avoids any ambiguity in the form of the corrections $W_\alpha$.

The steps that now follow are straightforward implementations of these prescriptions. At order $\alpha=2$, the right hand side $\mathcal{F}_2$ naturally belongs to $\mathbb{B}^C$. Hence, we can directly compute $W_2$ to use it for the next order.
At  order $\alpha=3$, however, the right hand side $\mathcal{F}_3$ contains terms in the critical sub-space $\mathbb{S}$. The solvability condition (\ref{SolCon}) demands that these terms must vanish. This imposition leads to the amplitude equations \eqref{AmplitudeLeft} and \eqref{AmplitudeRight}, which we fully analyze in the main text.

\end{document}